\documentclass{aa}
\usepackage[varg]{txfonts}

\usepackage{graphicx}
\usepackage{footnote}
\usepackage{txfonts}
\usepackage{amsmath,amssymb}
\usepackage{soul}
\usepackage{xcolor}
\usepackage{enumitem}
\usepackage[colorlinks=true, allcolors=blue]{hyperref}
\usepackage{bbm}
\usepackage{xspace}
\usepackage{tablefootnote}

\definecolor{BrickRed}{RGB}{182, 50, 28}
\definecolor{forestgreen}{rgb}{0.13, 0.55, 0.13}

\newcommand{\Yo}{\texttt{YOLO}\xspace}

\newcommand{\YoV}{\texttt{YOLOv3}\xspace}

\newcommand{\YoCL}{\texttt{YOLO-CL}\xspace}
\usepackage{lineno}
\begin{document}

\title{YOLO-CL cluster detection in  the Rubin/LSST DC2 simulations}
\titlerunning{YOLO-CL for LSST}
\authorrunning{Grishin et al.}
\author{
   Kirill Grishin\inst{1},
   Simona Mei\inst{1,2},
   Stephane Ilic\inst{3},
   Michel Aguena\inst{1},
   Dominique Boutigny\inst{4},
   Marie Paturel\inst{4}, 
   and the LSST Dark Energy Science Collaboration
}

 \institute{Universit\'e Paris Cit\'e, CNRS(/IN2P3), Astroparticule et Cosmologie, F-75013 Paris, France \email{grishin@apc.in2p3.fr, mei@apc.in2p3.fr}
\and Jet Propulsion Laboratory and Cahill Center for Astronomy \& Astrophysics, California Institute of Technology, 4800 Oak Grove Drive, Pasadena, California 91011, USA
\and IJCLab, Université Paris-Saclay, CNRS/IN2P3, IJCLab, 91405 Orsay, France 
\and LAPP, Université Savoie Mont Blanc, CNRS/IN2P3, Annecy; France
   }

\date{}
\abstract{The next generation large ground-based telescopes like the Vera Rubin Telescope Legacy Survey of Space and Time (LSST) and space missions like Euclid and the Nancy Roman Space Telescope will deliver wide area imaging surveys at unprecedented depth. In particular, LSST will provide galaxy cluster catalogs up to z$\sim$1 that can be used to constrain cosmological models once their selection function is well-understood.  Machine learning based cluster detection algorithms can be applied directly on images to circumvent systematics due to models, and photometric and photometric redshift catalogs.  In this work, we have applied the deep convolutional network YOLO for CLuster detection (\YoCL) to LSST simulations from the Dark Energy Science Collaboration Data Challenge 2 (DC2), and characterized the LSST \YoCL cluster selection function. We have trained and validated the network on images from a hybrid sample of (1) clusters observed in the Sloan Digital Sky Survey and detected with the red-sequence Matched-filter Probabilistic Percolation, and (2) dark matter haloes with masses $M_{200c} > 10^{14} M_{\odot}$ from the DC2 simulation, resampled to the SDSS resolution. We quantify the completeness and purity of the \YoCL cluster catalog with respect to DC2 haloes with $M_{200c} > 10^{14} M_{\odot}$. The \YoCL cluster catalog is 100\% and 94\% complete for halo mass $M_{200c} > 10^{14.6} M_{\odot}$ at $0.2<z<0.8$, and $M_{200c} > 10^{14} M_{\odot}$ and redshift $z \lesssim 1$,  respectively, with only 6\% false positive detections. We find that all the false positive detections are dark matter haloes with $ 10^{13.4} M_{\odot} \lesssim M_{200c} \lesssim 10^{14} M_{\odot}$, which corresponds to galaxy groups.  We also found that the \YoCL selection function is almost flat with respect to the halo mass at $0.2 \lesssim z \lesssim 0.9$. The overall performance of \YoCL is comparable or better than other cluster detection methods used for current and future optical and infrared surveys. \YoCL  shows better completeness for low mass clusters when compared to current detections based on Matched Filter cluster finding algorithms applied to Stage 3 surveys using the Sunyaev Zel'dovich effect, such as SPT-3G, and detects clusters at higher redshifts than X-ray-based catalogs. Future complementary cluster catalogs detected with the Sunyaev Zel'dovich effect will reach similar mass depth and will be directly comparable with optical cluster detections in LSST, providing cluster catalogs with unprecedented coverage in area, redshift and cluster properties. The strong advantage of \YoCL over traditional galaxy cluster detection techniques is that it works directly on images and does not require photometric and photometric redshift catalogs, nor does it need to mask stellar sources and artifacts.}


\maketitle

\section{Introduction} \label{sec:Intro}
Galaxy clusters are the largest gravitationally bound structures in the Universe, and their distribution is a probe for cosmological models. Upcoming deep large-scale survey like those performed with the Vera C. Rubin Observatory~\citep{2018cosp...42E1651K}, the Euclid space telescope~\citep{Euclid-r} and the Nancy Grace Roman Space Telescope~\citep{2021MNRAS.507.1746E} will give us unprecedented deep optical and infrared imaging of hundreds of thousands of clusters up to z$\sim$2. 

In particular, the Vera Rubin Telescope Legacy Survey of Space and Time (LSST; \citealp{2009arXiv0912.0201L, 2019ApJ...873..111I})  will deliver deep optical imaging data over $\sim$20,000 sq. deg. of the sky. LSST will observe in six bandpasses (u, g, r, i, z, y) and reach a depth of r$\sim$27.5mag on about half of the sky~\citep{2012SPIE.8446E..6BO, 2006SPIE.6273E..0YO}. These observations  will permit us to obtain constraints on cosmological models using galaxy clusters, once we can provide a precise selection function.  

Cluster detection in optical and near-infrared multi-wavelength imaging surveys is mainly based on the search of spatial overdensities of galaxies of a given class, which can be quiescent, line-emitter, massive, etc. \citep[e.g., ][]{2005ApJS..157....1G, 2007ApJ...660..239K, 2009ApJ...697.1842K, 2009ApJS..183..197W, 2012ApJS..199...34W, 2010MNRAS.404.1551S, 2010ApJS..191..254H,  2011ApJ...736...21S, 2012ApJ...746..188M, 2011ApJS..193....8B, 2013ApJ...769...79W, 2014ApJ...786...17W, 2014ApJ...785..104R, 
2015ApJ...804..117M,2016MNRAS.455.3020L, 2016ApJ...829...44L, 2019MNRAS.485..498M, 2023MNRAS.519.2630W}.
Most of these methods require a high-quality photometric calibration, an accurate calibration of galaxy colors as a function of redshift, and unbiased photometric and photometric redshift catalogs. Photometric catalogs might be affected by aperture or model choices in measuring magnitudes and background subtraction. These systematics propagate to the estimation of photometric redshifts, which also rely on being calibrated on available spectroscopic redshift samples and galaxy spectral energy distribution templates that do not cover the entire galaxy population \citep[e.g., ][]{2024ApJ...967L...6M}. These uncertainties on both photometric and photometric redshift catalogs make it essential to complement traditional cluster detection algorithms with new techniques that do not rely on catalogs, but instead work directly on images, such as deep machine learning (ML) neural network.

Over the last years, deep ML techniques were widely used in astrophysics for different purposes~\citep{2023PASA...40....1H}, including object classification~\citep{2018MNRAS.476.3661D, 2023A&A...676A..40A}, estimation of redshift of individual galaxies~\citep{2019A&A...621A..26P, 2021MNRAS.505.4847H}, solution of ill-posed problems, including reconstructions of matter distributions~\citep{2020MNRAS.492.5023J, 2022arXiv220105571C, 2023MNRAS.523.6272C}. The purity of the samples, defined as the percentage of true objects recovered by the network as opposite to false detections, was high enough to search for rare or elusive objects \citep{2017Natur.548..555H,2021A&A...647A.116C}. Among these methods, convolutional neural networks (CNN) are well adapted for object detection and characterization in astrophysics \citep[e.g.,] []{2015ApJS..221....8H,2018ApJ...858..114H,2018MNRAS.478.5410D, 2019A&A...621A..26P,2021MNRAS.501.4359Z,2022A&A...657A..90E,2022A&A...665A..34D, 2023MNRAS.520.3529E, 2023A&A...671A..99E}, in particular for galaxy cluster detection \citep[e.g.,] []{2019MNRAS.490.5770C,2020A&A...634A..81B,2021A&A...653A.106H, 2021MNRAS.507.4149L, 2023A&A...677A.101G}.  

Recently, our team developed a cluster detection method modifying the well-known detection-oriented deep machine learning neural network ``\texttt{You only look once}" \citep[\Yo,][]{2015arXiv150602640R,2016arXiv161208242R}. Our network, \YoCL \citep[YOLO for CLuster detection; ][]{2023A&A...677A.101G}, detects galaxy clusters on multi-wavelength images, and shows a higher performance with respect to traditional cluster detection algorithms in obtaining cluster catalogs with high completeness and purity. When applied to the Sloan Digital Sky Survey \citep[SDSS; ][]{2000AJ....120.1579Y},   \YoCL provides cluster catalogs that are complete at $\sim 98\%$ for X-ray detected clusters with $ {\rm I_{X, 500}} \gtrsim 20 \times 10^{-15} \ {\rm erg/s/cm^2/arcmin^2}$  at $0.2 \lesssim z \lesssim 0.6$, and of $\sim 100\%$ for clusters with $ {\rm I_{X, 500}} \gtrsim 30 \times 10^{-15} \ {\rm erg/s/cm^2/arcmin^2}$ at $ 0.3 \lesssim z \lesssim  0.6$. The contamination from false detections is $\sim 2\%$.
It is also interesting that \citet{2023A&A...677A.101G} found the \YoCL selection function is flat as a function of redshift, with respect to the X-ray mean surface brightness. The advantage of \YoCL, and other ML networks that work directly on images, is that  they are independent of models and systematics that might arise when building photometric and photometric redshift catalogs in traditional methods. They also do not need stellar sources and artifacts to be masked. If the training sample is representative of the entire observed sample, the ML methods should be less impacted by modeling choices and systematics.

In this paper, we evaluate the \YoCL efficiency in detecting galaxy clusters in the LSST survey. Given that LSST observations did not  start yet, we apply the network on simulations from the LSST Data Challenge 2 \citep[DC2; ][]{2021ApJS..253...31L}, which were developed within the LSST Dark Energy Science Collaboration (DESC\footnote{\url{https://lsstdesc.org/}}). We quantify the \YoCL cluster catalog selection function in terms of completeness and purity (see below) with respect to DC2 haloes with $M_{200c} > 10^{14} M_{\odot}$. The \YoCL cluster catalog is 100\% and 94\% complete for halo mass $M_{200c} > 10^{14.6} M_{\odot}$ at $0.2<z<0.8$, and 94\% complete for $M_{200c} > 10^{14} M_{\odot}$ and redshift $z \lesssim 1$,  respectively, with only 6\% false positive detections. This contamination is expected from the intrinsic accuracy of convolutional neural networks, and our network is highly efficient with respect to traditional cluster detection algorithms based on photometric and photometric redshift catalogs. It is interesting that all the false positive detections are groups with $ 10^{13.4} M_{\odot} \lesssim M_{200c} \lesssim 10^{14} M_{\odot}$, and that the catalog selection function is flat with respect to the halo mass at $0.2 \lesssim z \lesssim 0.9$.

This article is organized as it follows: in Section~\ref{sec:dc2} we describe the observations and simulations used to train and validate our network. In Section~\ref{sec:yolocl} we present \YoCL and its training and validation.  The results and the discussion and conclusions are presented in Section~\ref{sec:results} and Section~\ref{sec:discuss}, respectively. The summary is in Sec.~\ref{sec:summary}. All magnitudes are given in the AB system~\citep{1983ApJ...266..713O, 2005PASP..117.1049S}. We adopt a $\Lambda CDM$ cosmology, with $\Omega_M  =0.3$,
 $\Omega_{\Lambda} =0.7$, $h=0.72$, and $\sigma_8 = 0.8$.

\section{Observations and simulations}

Since the DESC DC2 simulated area includes only $\approx 2,000$ synthetic galaxy clusters (see Sec.~\ref{sec:dc2}) and we need at least 10,000 objects for training our network, we trained \YoCL on a hybrid sample of cluster images that includes both the same set of SDSS observed images \citep{2009ApJS..182..543A} 
that we used in \citet{2023A&A...677A.101G}, and   synthetic cluster images from the DESC DC2 simulations.

This strategy is widely used in astrophysics when the target sample (in our case the LSST DC2 simulations) is large enough to provide a statistical application of a network, but too small to be used for the network training and validation. 
In the case of convolutional networks, such as \YoCL,  \citet{2018MNRAS.476.3661D} demonstrated that transfer learning allows for rapid adaptation from one astrophysical survey application to another. Specifically, the weights obtained by training a convolutional network on images from a given survey can be efficiently transferred to another survey by fine-tuning them, i.e., by retraining the network adding a smaller number of images from the new survey, roughly an order of magnitude fewer than the initial training sample. In their case, the initial survey was SDSS, and they applied transfer learning to the Dark Energy Survey \citep{2018ApJS..239...18A}. We demonstrate in this section that this approach is also effective when re-training \YoCL using our initial training set from SDSS as utilized in \citet{2023A&A...677A.101G}, and incorporating approximately one order of magnitude fewer synthetic cluster images from the DESC DC2 simulations.


\subsection{The SDSS observations}
\label{sec:sdss}

 The SDSS is an imaging survey that was performed with the 2.5-m. Apache Point telescope in five optical bandpasses (u, g, r, i, z) using the SDSS camera in a scanning regime. It covers $\sim$ 14,055 sq. deg. of the sky in two main areas in the Northern hemisphere split by the Milky Way: one within 7h < RA < 16h and -1 deg <Dec< +62 deg. and the other within 20h<RA<2h and -11 deg.<Dec<+35 deg. The 5-$\sigma$ point-source depth in the g, r and i bandpasses is 23.13, 22.70 and 22.20 mag, respectively. The seeing quality for SDSS images varies from 1.2 to 2.0~arcsec\footnote{\url{https://www.sdss4.org/dr17/imaging/other_info/}}.

As reference SDSS cluster catalog, we used the red-sequence Matched-filter Probabilistic Percolation
(redMaPPer) Data Release 8 (DR8) catalog from \citet{2014ApJ...785..104R}. The redMaPPer algorithm finds overdensities of red sequence galaxies in large photometric surveys. The cluster catalog that we used\footnote{Version 6.3 of the catalog, from \url{risa.stanford.edu/redMaPPer}.} covers  $\sim$10,000 square degrees of the SDSS DR8 data release, and includes 26,111 clusters over the redshift range z $\in$ [0.08, 0.55].  The redMaPPer catalog is 100\% complete up to $z=0.35$ for clusters from the MCXC (Meta-Catalog of X-Ray Detected Clusters of Galaxies) X-ray detection catalog \citep{2011A&A...534A.109P}, with temperature $T_X \gtrsim 3.5 keV$, and luminosity $L_X \gtrsim 2 \times 10^{44}$ erg  s$^{-1}$, decreasing to 90\% completeness at $L_X \sim 10^{43}$ erg \ s$^{-1}$. The centers of 86\% of the redMaPPer clusters correspond well to their X-ray centers \citep{2014ApJ...785..104R}. For each cluster, redMaPPer provides its position,  the richness $\lambda$\footnote{By definition, the cluster richness is the number of cluster members above a given luminosity. For redMaPPer it is defined as a sum of the probability of being a cluster member over all galaxies in a cluster field~\citep{2009ApJ...703..601R}.}, and a list of cluster members. The richness is correlated to the cluster mass. All redMaPPer rich clusters ($\lambda > 100$) are detected in the X-ray ROSAT All Sky Survey \citep{1999A&A...349..389V}.
  
 We excluded clusters with redshifts $z<0.2$ from the original redMaPPer cluster catalog, because they cover regions in the sky larger than the images that we consider when optimizing our network execution time and computational power (see sec. 3.2). Our final redMaPPer catalog includes 24,406 clusters, whose distribution is shown in Fig.~1 from \citet{2023A&A...677A.101G}. 

For the network training and validation, we used JPEG color images of the original SDSS DR16 images centered on each of the 24,406 redMaPPer clusters,  using the \texttt{ImgCutout} web service\footnote{\url{http://skyserver.sdss.org/dr16/en/help/docs/api.aspx\#imgcutout}}. These images were derived from the \textit{g}, \textit{r}, and \textit{i}-band FITS corrected frame files from the Science Archive Server, and the color images are built using the conversion algorithm\footnote{Detailed here: \url{https://www.sdss.org/dr16/imaging/jpg-images-on-skyserver}}
based on \citet{2004PASP..116..133L}.  We chose these three bandpasses because they are sufficient to identify passive early-type galaxies in clusters at $z \lesssim 1$.

\subsection{The DESC DC2 simulation}
\label{sec:dc2}

In ten years, LSST will reach the 5-$\sigma$ point-source depth of  27.4, 27.5, and 26.8~mag in the g, r and i bandpasses, respectively~\citep{2019ApJ...873..111I}. This will allow to build a catalog of 20 billion individual galaxies, and over 100,000 galaxy clusters at z<1.2. The average seeing quality at the Rubin telescope site is 0.67'' with a best value of 0.4'', which is very close to the best spatial resolution that can be achieved from the ground. 

 The primary goal of the LSST DESC DC2 simulation is to create realistic LSST synthetic observations that can be used to test all DESC primary pipelines. 
DC2 is based on the Outer Rim cosmological N-body simulation, that contains around a trillion particles in 4.225 Gpc$^3$ of co-moving volume \citep{2019ApJS..245...16H}. An extragalactic catalog, CosmoDC2, was built from the snapshots of Outer Rim simulation by: 1) assigning galaxies to each halo of the dark matter simulation with properties obtained from empirical relations ~\citep{2019MNRAS.488.3143B}, and 2) fully characterizing galaxies in this sample adding missing properties derived from the semi-empirical model (SAM) Galacticus~\citep{2012NewA...17..175B}.

The CosmoDC2 catalog was used to simulate images over an area of 445~sq.~deg., with galaxies at z<3. The sample of galaxies in the initial truth catalog is complete down to r=28.0~mag, and galaxies fainter than r=29.0~mag are excluded from the simulations for computation performance purposes.  
The catalog is stored in the HEALPix format~\citep{2005ApJ...622..759G} and split into three redshift bins: 0<z<1, 1<z<2 and 2<z<3. The quality of this catalog was evaluated in the framework of the LSST DESC collaboration using the DESCQA validation framework~\citep{2018ApJS..234...36M}. 
This validation confirmed that the  simulation reproduce reasonably well galaxies, their properties, and their distribution in the Universe~\citep{2022OJAp....5E...1K}. This makes the  DC2 simulation one of the best dataset to test the DESC cosmological pipelines and algorithms, including cluster finders.

The simulation includes both a catalog and synthetic images. The simulation of the DC2 synthetic images consisted of two mains steps: 1) simulation of raw images that resemble those obtained with LSSTCam, and 2) reduction of these raw images using the LSST science pipeline~\footnote{\url{https://pipelines.lsst.io/}}, based on the Hyper Supreme-Cam pipeline \citep{2018PASJ...70S...5B}. On the first step, each object from cosmoDC2 catalog was simulated using the GalSim package~\citep{2015A&C....10..121R}, taking into account the LSST depth and noise, accounting for CCD effects, night sky background~\citep{2016SPIE.9910E..1AY}, cosmic ray hits, etc.  Galaxy colors and spectral energy distributions were modeled using templates from~\citet{2003MNRAS.344.1000B}. 

The raw synthetic images were then processed by the LSST science pipeline, which covers: 1) single-frame processing, by basic corrections like bias subtraction, non-linearity and flat-field corrections, and first iteration of astrometric and photometric calibration 2) joint calibration, which uses synthetic observations of the same area of the sky from different frames to improve the calibration 3) image co-addition, when individual images are resampled on the same coordinate grid, and then coadded, and 4) source detection. The 5$\sigma$ point-source depth of the simulation in the r-band is 27.3 mag, which corresponds to 5 years of the LSST survey, the deeper DC2 images on a large area currently available. 

Using the Dark Energy Survey \citep[DES; ][]{2018ApJS..239...18A} exposure checker~\citep{2016A&C....16...99M}, a few dozens of DESC members performed a quality check of $\sim$9,000 synthetic co-added images, which did not show substantial issues \citep{2021ApJS..253...31L}. The galaxy catalogs comply with the LSST Science Requirements~\citep{LPM-17} and the DESC Science Requirements~\citep{2018arXiv180901669T}.
These images are expected to have properties, including depth and seeing quality, very close to those that will be obtained with LSSTCam \citep{2018SPIE10705E..0DR}. 

\begin{figure*}[ht!]
    \includegraphics[width=2\columnwidth]{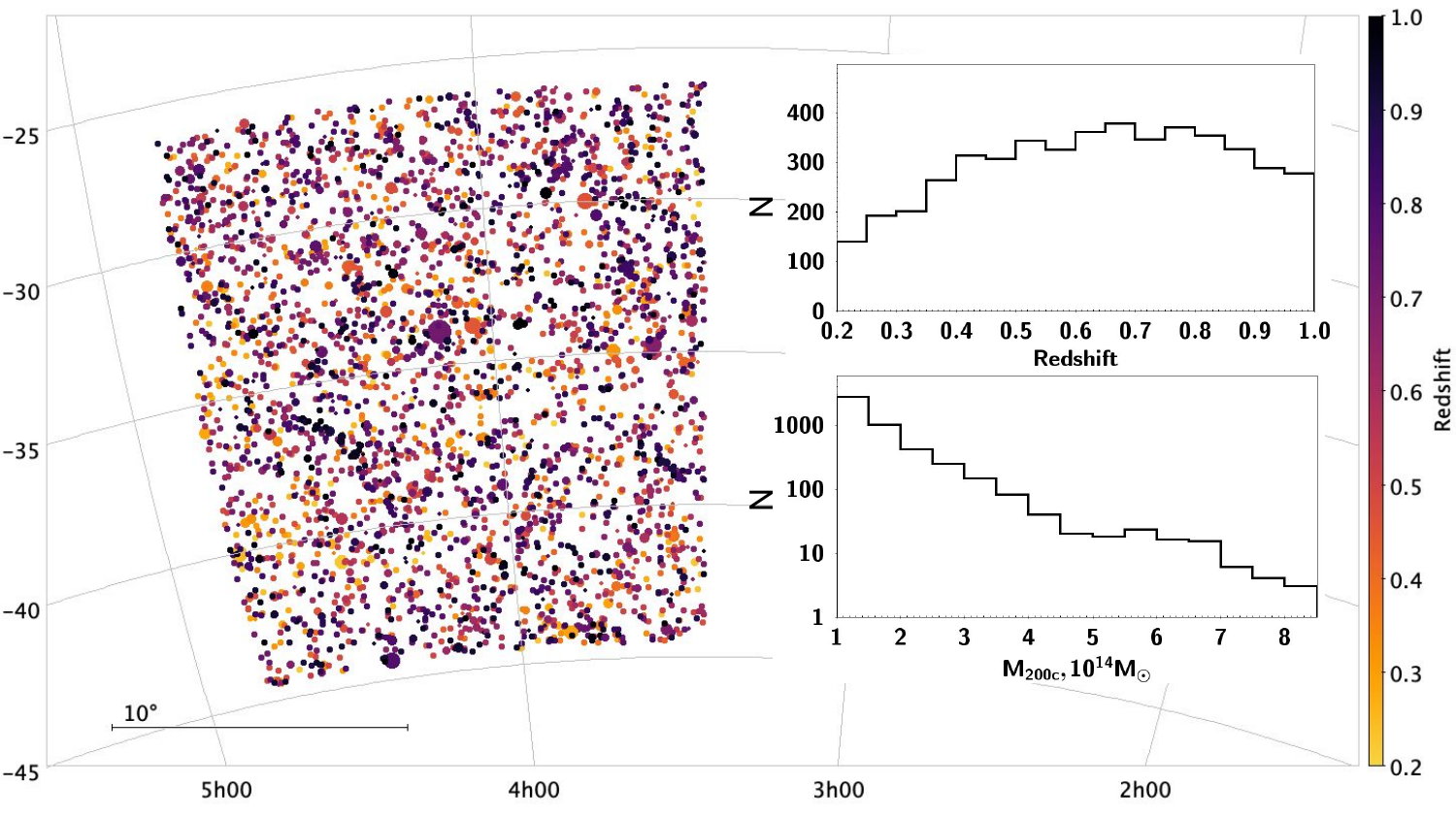}
    \caption{Sky map with the positions of the 2,342 total CosmoDC2 clusters with M$_{200c}>10^{14}M_{\odot}$ that we used for the \YoCL training and validation. Larger circle sizes indicate larger masses, and redshift is coded by color, as indicated in the right bar). In the insert: the dark matter halo redshift and mass distributions.}
    \label{fig:map_dc2}
\end{figure*}

The cosmoDC2 v1.1.4 catalog includes 2,342 dark matter halos with 0.2<z<1 and M$_{200c}$ > 10$^{14}~M_{\odot}$\footnote{M$_{200c}$ is defined as the mass within the circular region of radius R$_{200}$ containing a mean mass density equal to two hundred times the critical density of the Universe at a given redshift.} (the typical minimal halo mass of virialization from \citealp{2008ApJ...672..122E} that defines galaxy clusters, hereafter we will refer to these haloes as DC2 clusters) and redshift in the range 0.2<z<1.0. Hereafter, we refer to this sample as our DC2 "true cluster" sample.
We exclude halos on the simulation edges, which are not entirely included in the images. In this work, we use this sample as the "true cluster" sample. Fig.~\ref{fig:map_dc2} shows our DC2 cluster sample and its redshift and mass distributions. 

For each halo, the catalog includes its position, the true redshift, the dark matter halo mass M$_{200c}$, and a richness parameter defined as the sum of the probabilities for galaxies brighter than $m^*(z)+2$ to be a halo member. Here $ m^* $ is the characteristic magnitude that corresponds to the luminosity of the knee of the Press–Schechter luminosity function \citep{1974ApJ...187..425P} at the redshift of the cluster.  To find $ m^* $, we fitted the galaxy luminosity function in the K-band~\citep{2006ApJ...650L..99L}. Then, we predict $ m^* $ in optical bands using the PEGASE2 library~\citep{1999astro.ph.12179F} for a burst galaxy that passively evolves from z=3. The probability for the galaxy to be a cluster member was computed assigning a weight depending on the projected distance from the cluster center following~\citet{2012ApJ...746..178R,2014ApJ...785..104R}.

To generate composite color images, we used the {\sc deepCoadd} frames delivered by the LSST pipeline in the DC2 Run2.2 simulation run \citep{2021ApJS..253...31L} for the cosmoDC2 v1.1.4 extragalactic catalog \citep{2019ApJS..245...26K}.  These images are fully reduced, calibrated, sky subtracted and co-added science frames with a pixel scale of 0.2''/pix. To make our analysis fully consistent with the SDSS images, we have resampled the DC2 images to the SDSS pixel scale of 0.39''/pix using the astropy-based {\sc reproject} package \citep{2020ascl.soft11023R}.

To build composite JPEG color images for DC2 simulation we used the same algorithm used in the SDSS survey \citep{2001ASPC..238..269L}. This algorithm has two main parameters: nonlinearity (Q) and flux scale ($\alpha$). For SDSS, the parameters are Q=8 and $\alpha$=0.2\footnote{\url{https://sdss4.org/dr17/imaging/jpg-images-on-skyserver/}}. For the DC2 color images, we used Q=8 and $\alpha$=0.08,  in order to partially compensate the depth and magnitude zeropoint difference (the zero magnitudes are $m_0^{SDSS}=22.5$~mag and $m_0^{LSST}=27$~mag). In fact, with $\alpha$=0.08, the DC2 scale visually reproduces the SDSS scale. We also adjusted the DC2 flux count range to have a similar range in surface brightness as in SDSS. We performed a sky subtraction, and registered the composite images on a final JPEG scale from 0 to 255. We set to zero and 255 all pixels with fluxes less than zero and larger than 255, respectively.




\section{\YoCL training and validation}
\label{sec:yolocl}

\subsection{\YoCL}

\YoCL\footnote{GITHUB PAGE} is based on the the third iteration
of \Yo, \YoV \citep{2018arXiv180402767R}, which represents
a significant improvement over the first  versions, and proved to be very well adapted for cluster detection~\citep{2023A&A...677A.101G}. We outline here the algorithm main characterists, and more details can be found in \citet{2023A&A...677A.101G}. The \Yo architecture applies a single neural network to images, combining object detection and classification into a single process. This results in several orders of magnitude faster execution times, compared to other detection convolutional networks such as R-CNN \citep[Region Based Convolutional Neural Networks,][and the following developments Fast and Faster R-CNN]{2013arXiv1311.2524G}. 

The network divides the image into a $S{\times}\,S$ grid of cells, within which the detection and classification are performed. For each object detection the network predicts $B$ bounding boxes, to which it assigns a set of parameters, including its position, size, the probability of being an object and the probability of belonging to a certain class of objects. The network is trained on a sample of images on which it optimizes the parameters to better detect and classify objects (i.e.,  converges on the optimal weights). 

During the training process \YoCL optimizes a multi-component loss function $\mathcal{L}$ \citep{2015arXiv150602640R, 2023A&A...677A.101G}: 
\begin{equation}
    \label{eq:yolo_loss}
    \mathcal{L} =  \mathcal{L}_{\rm obj} + \mathcal{L}_{\rm bbox} +\mathcal{L}_{\rm class} \ .
\end{equation}
Where $\mathcal{L}_{\rm obj}$ is the "objectness loss" and optimizes the object identification, $\mathcal{L}_{\rm bbox}$ is the "bounding box loss" and optimizes the bounding box position and size, and $\mathcal{L}_{\rm class}$ is the ``classification loss'' and optimizes the object class. The loss functions quantify the distance between the true parameter values and those estimated by the network. With respect to the original \Yo `classification loss'' function that considers several object classes, in \YoCL we removed multiple object classes because we use a single object class, which is "cluster". As "bounding box loss", we used the generalized Intersection over Union
(gIoU) loss ~\citep{Rezatofighi_2018_CVPR}. In fact, the traditional IoU (Intersection over Union\footnote{The IoU is defined as the ratio between the area of intersection and the area of union between the detected object bounding box and the "true object" bounding box \citep{2015arXiv150602640R}}) metric does not permit us to optimize the corresponding loss term when the true and predicted bounding boxes are non-overlapping. More details can be found in \citet{2023A&A...677A.101G}.

The \YoCL training consists of several iterations, which are called epochs. At each epoch all the images from the training sample are an input for the network which optimizes the network weights and bias that decrease the loss function, making the distance between the true values and those estimated by the network closer. 
The network is then validated on a validation sample.

The final network output is a catalog of detections with an associated detection probability (see below).

\begin{table*}[ht!]
   \centering
    \caption{Settings used for the \YoCL training}
    \resizebox{!}{0.8cm}{
    \begin{tabular}{ c c c c c c }
    \hline
    Image resolution & Batch size & Number of training & Data augmentation & Augmentation frequency & gIoU threshold\\
     &  & epochs & technique &   per technique& \\[0.1cm]
    \hline
    $1024\times1024$ & 2 & 100 & horizontal flip, vertical flip, transpose, translate & 50 \% & 50 \% \\
    $512\times512$ & 8 & 100 & horizontal flip, vertical flip, transpose, translate & 50 \% & 50 \% \\
    \hline
    \end{tabular}
    }
    \label{tab:settings}
\end{table*}

\begin{figure*}[ht!]
    \centering
    \includegraphics[width=0.49\textwidth]{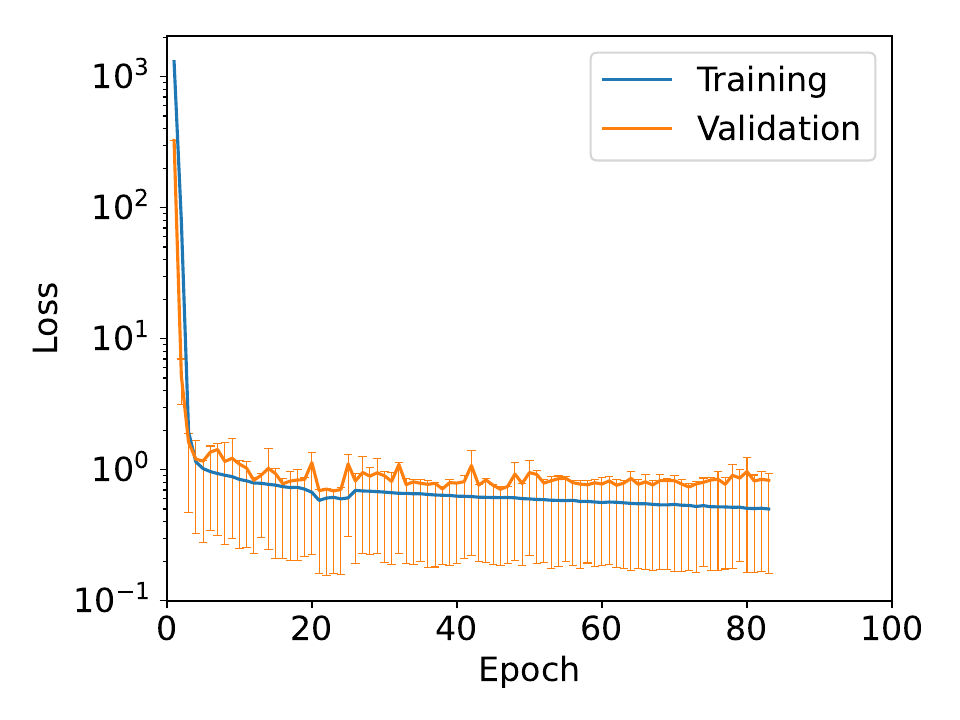}
    \includegraphics[width=0.49\textwidth]{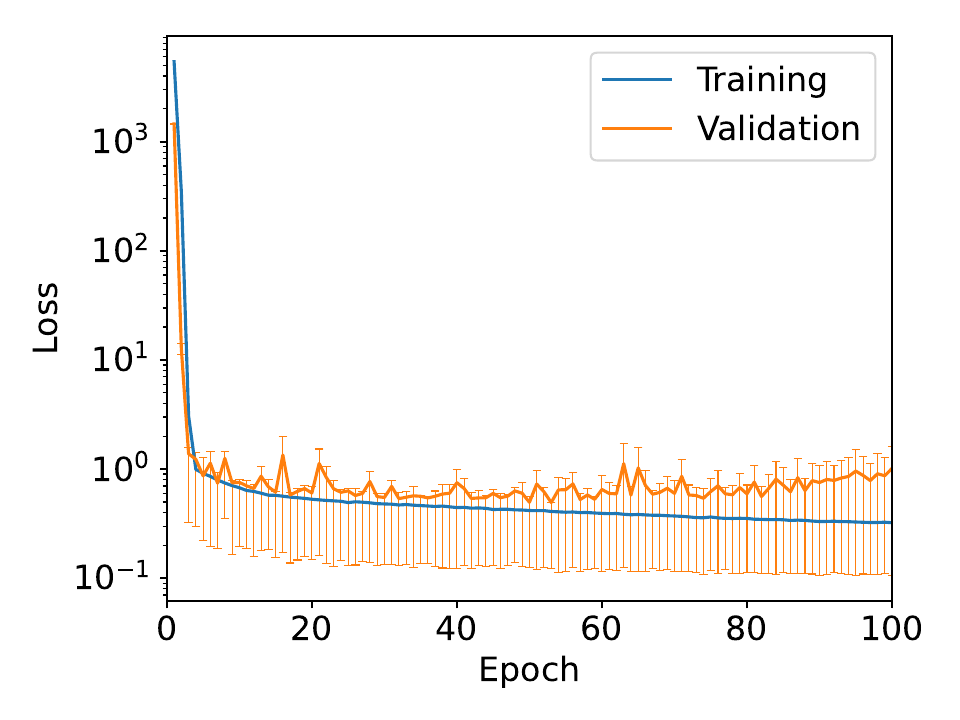}
    \caption{The \YoCL loss functions for the training (blue) and validation (orange) samples, for the 512x512 (left) and 1024x1024 (right) images. The vertical bars show the 1~$\sigma$ standard deviation of the validation loss. The training and validation loss functions converge in a smooth way, and their good agreement confirms the network stability in both cases. } \label{fig:loss_lowthr}
\end{figure*}

\subsection{Training and validation}
\label{sec:training}

We used two equal hybrid samples of 12,203 redMaPPer and 1,171 DC2 cluster images each for both training and validating \YoCL, with the same number but different images for the training and validation. Each of these two samples has identical redshift and mass distribution, for a total of 24,406 redMaPPer and 2,342 DC2 cluster images. Our hybrid training and validation sample approach makes the \YoCL learning invariant to the differences in object densities, and all the other differences between SDSS and DC2.  


Following \citet{2023A&A...677A.101G}, we start with images of  dimension   2048$\times$2048 pixels, which corresponds to $\sim$13.5 x 13.5 arcmin$^2$, twice the size of a typical cluster virial radius of 1~Mpc at $z\sim 0.2$, and much larger than the typical cluster virial radius at $z>0.5$. For the input to the first layer of the network, we resize each image by average pooling to 512$\times$512 pixels (with a pixel size equal to eight times the LSST resolution\footnote{four times the SDSS resolution}) and 1024$\times$1024 pixels (with a pixel size equal to the four time of the LSST resolution\footnote{the double of the SDSS resolution}), and keep the same stride parameters as in the original \YoV publication, namely 8, 16, and 32. 

These image sizes and stride parameters are a good compromise between keeping high image resolution and our computational power. Our  training and validation runs  were performed on Centre de Calcul IN2P3\footnote{\url{https://cc.in2p3.fr/}} computing cluster on a NVIDIA Tesla V100-SXM2-32GB GPU, equipped with 32 GB of memory.

\subsubsection{Hyperparameter optimization}

Our hyperparameters optimization is performed with respect to memory limits and the stability of the training. Since the weight optimization during the training is done using a gradient descent, the whole process can be a subject to instabilities. There are two main hyper-parameters responsible for the mitigation of these instabilities: the batch size and the learning rate.
The size of the training sample is too large to store in memory, and it is not possible to complete the training on the entire sample in one iteration. To overcome this limitation, we split our training sample in subsets (batches) that are processed by the network at the same time. The batch size is limited by two main factors: it cannot be too small, because in this case the derived direction of the gradient would be unstable, and at the same time it cannot be too big, given that memory resources are limited.  Due to memory limitations, we used a batch size of 8 for the 512x512 images and of 2 for the 1024x1024 images. 

The other hyper-parameter that is crucial for training is the learning rate. It defines how big the weight variations can be at each epoch. It cannot be too small, otherwise the most optimal weight configuration would never be achieved, and it cannot be too big because it would make the training process less stable.  We choose the learning rate varying with the epoch: it starts from a some small value and grows up during a few first epochs, called ``warm-up'' epochs, and after reaching its maximum values it asymptotically goes down to the final values \citep{2023A&A...677A.101G}. Starting, maximal and final values of the learning rate as well as the number of ``warm-up'' epochs are also hyper-parameters, and should be defined before the training. We start by setting a learning rate of 10$^{-10}$, which grows to 10$^{-5}$ during the first eight warm-up epochs, and then slowly decreases to 10$^{-6}$. 

Our input image cutouts are centered on the redMaPPer cluster or DC2 selected dark matter halo positions. This centering should not have an impact on the network learning, which should understand that cluster features should not depend on its position in the image. For this reason, we apply data augmentation, including translation and flipping of a random quantity between zero and half of the image, which change the initial cluster position in the image. This forces the network to focus on the relevant features associated with clusters, independently of their position in the images.

We provide the main parameters of the training configuration in Table~\ref{tab:settings}.

\begin{figure*}[ht!]
    \centering
    \includegraphics[width=0.49\textwidth]{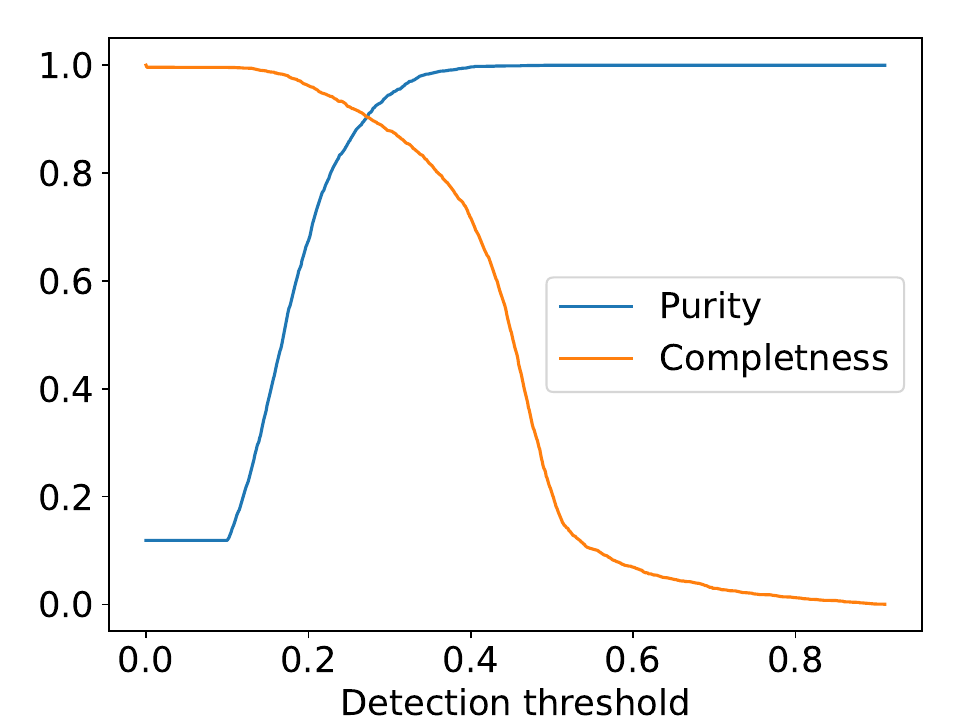}
    \includegraphics[width=0.49\textwidth]{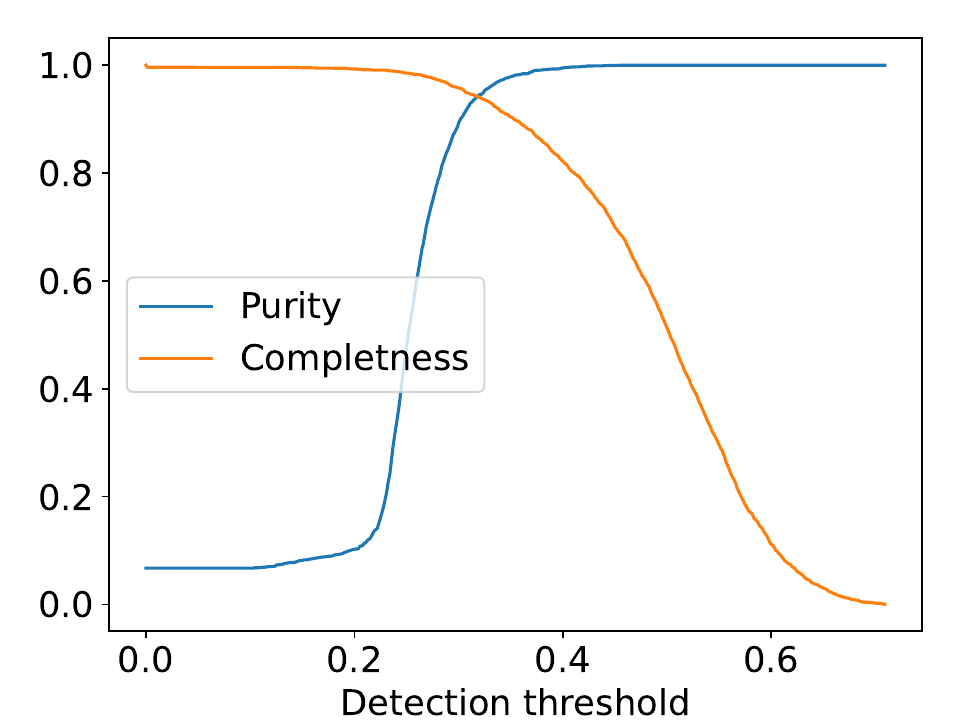}
    \caption{The purity and completeness of the \YoCL DC2 detection catalogs for 512x512 (left) and 1024x1024 (right) images as a function of the detection threshold. The best purity and completeness are 90\% and 94\% for the 512x512 and 1024x1024 pixel images, respectively.
    } \label{fig:pc_lowthr}
\end{figure*}

\begin{figure*}[ht!]
    \centering
    \includegraphics[width=0.49\textwidth]{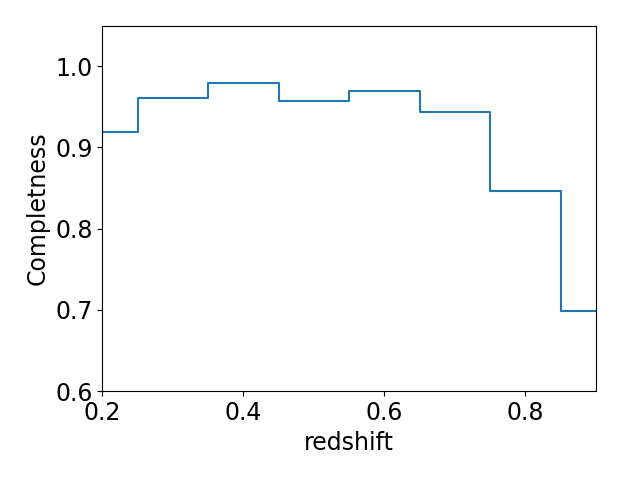}
    \includegraphics[width=0.49\textwidth]{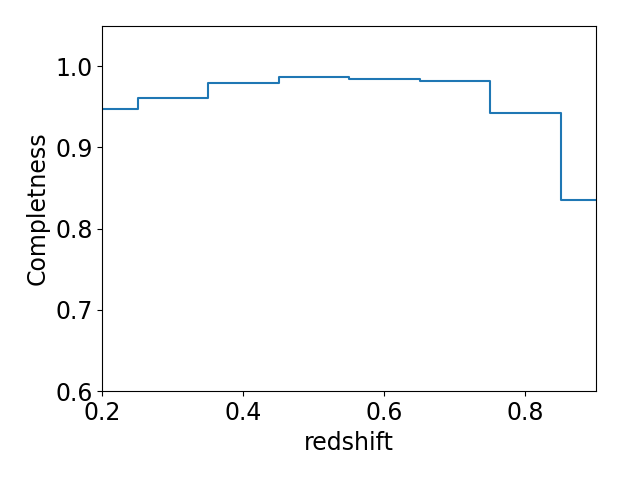}\\
    \includegraphics[width=0.49\textwidth]{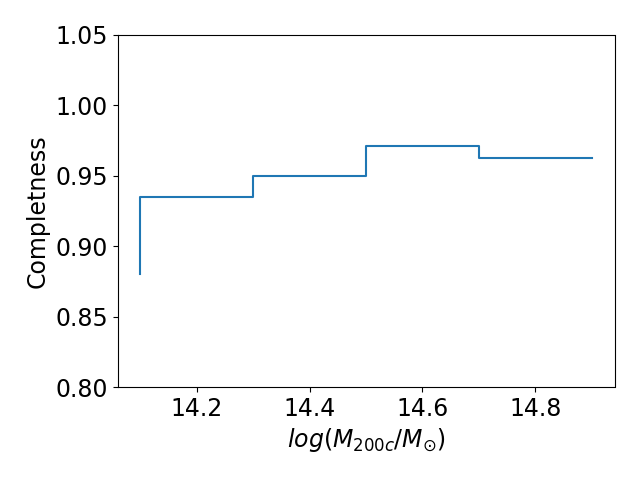}
    \includegraphics[width=0.49\textwidth]{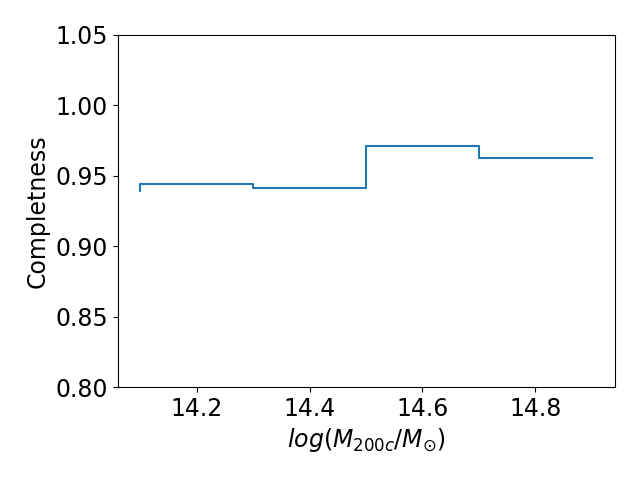}
    \caption{The \YoCL DC2 detection completeness as a function of redshift (Top), and halo mass M$_{200c}$ (Bottom) for the 512x512 (left) and 1024x1024 (right) pixel images. The completeness as a function of redshift is almost flat at the region 0.2<z<0.8 with an average value of 0.85 for 512x512 and 0.95 for 1024x1024. } \label{fig:pc_massz}
\end{figure*}

\begin{figure*}[ht!]
    \centering
    \includegraphics[width=0.49\linewidth]{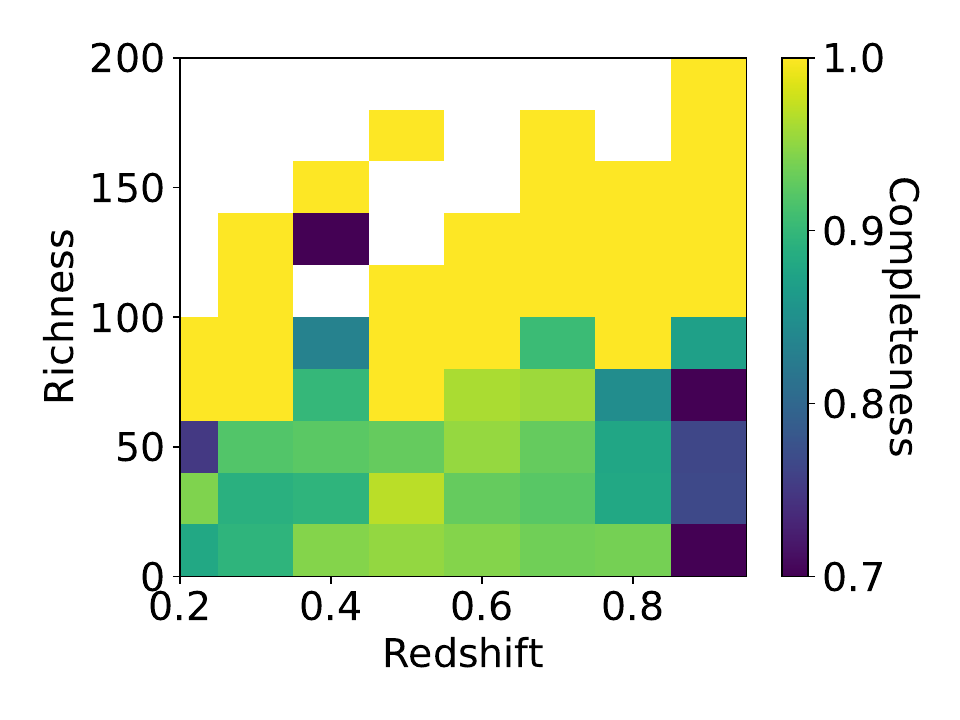}
    \includegraphics[width=0.49\linewidth]{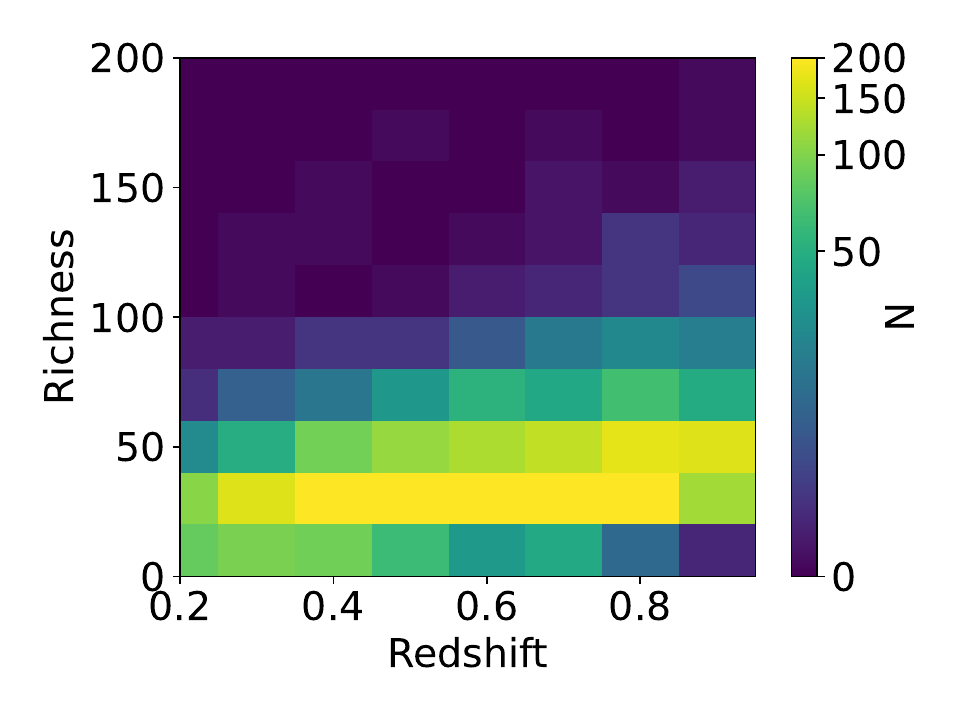}
    \includegraphics[width=0.49\linewidth]{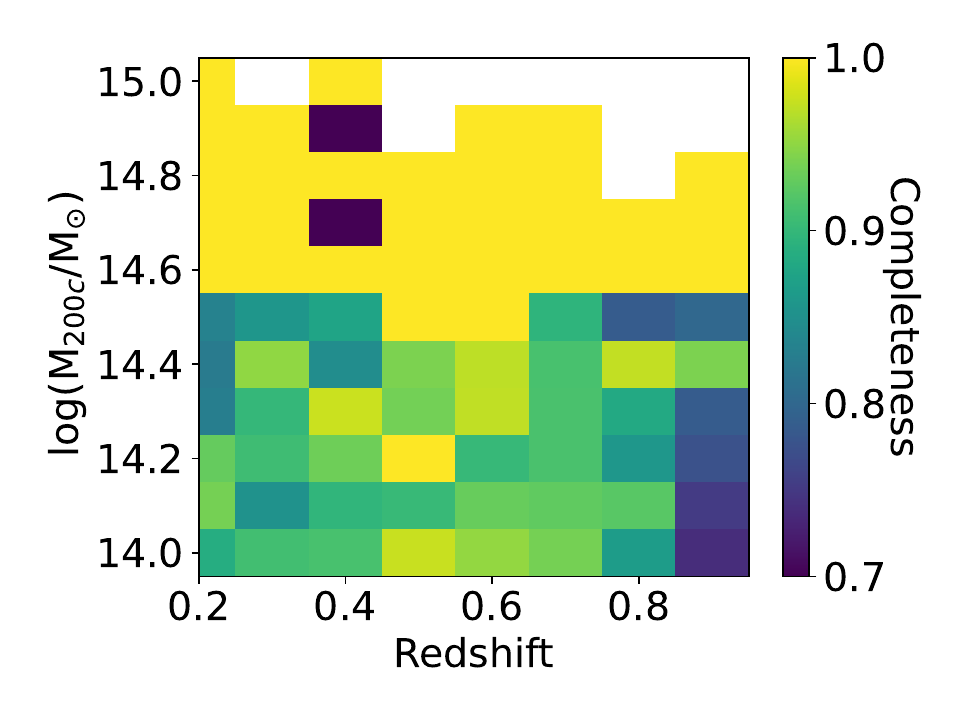}
    \includegraphics[width=0.49\linewidth]{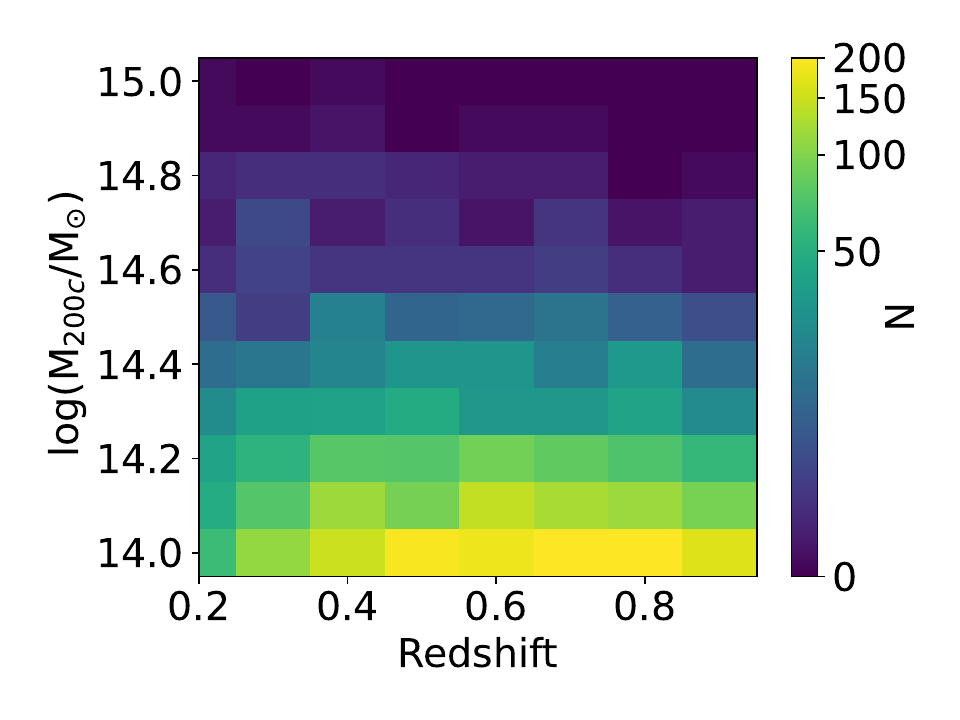}
    \caption{Left: \YoCL DC2 detection completeness as a function of redshift and richness (top) or halo mass (bottom). Right: the number of halos as a function of redshift and richness (top) or halo mass (bottom). The colored vertical bars show the color scale for the completeness (left) and the number of haloes (right).
    The \YoCL selection is almost flat as a function of redshift up to $z\sim0.8$ when we consider the halo mass. The catalog is $\sim$100\% complete for $M_{200c} \gtrsim 10^{14.6}M_{\odot}$ and richness $\gtrsim 100$ at all redshifts. When characterizing halos by their richness, the completeness is less flat as a function of redshift, as also shown for SDSS observations in \citet{2023A&A...677A.101G}, and the completeness decreases abruptly to $\sim 70-75\%$ at $z>0.8$ and $M_{200c} \lesssim 10^{14}M_{\odot}$. Comparing the figures on the left and on the right, some bins show very low completeness on the left only because there are no clusters in those bins. 
} \label{fig:c_zlam}
\end{figure*}

\begin{figure*}[ht!]
    \centering
    \includegraphics[width=0.49\linewidth]{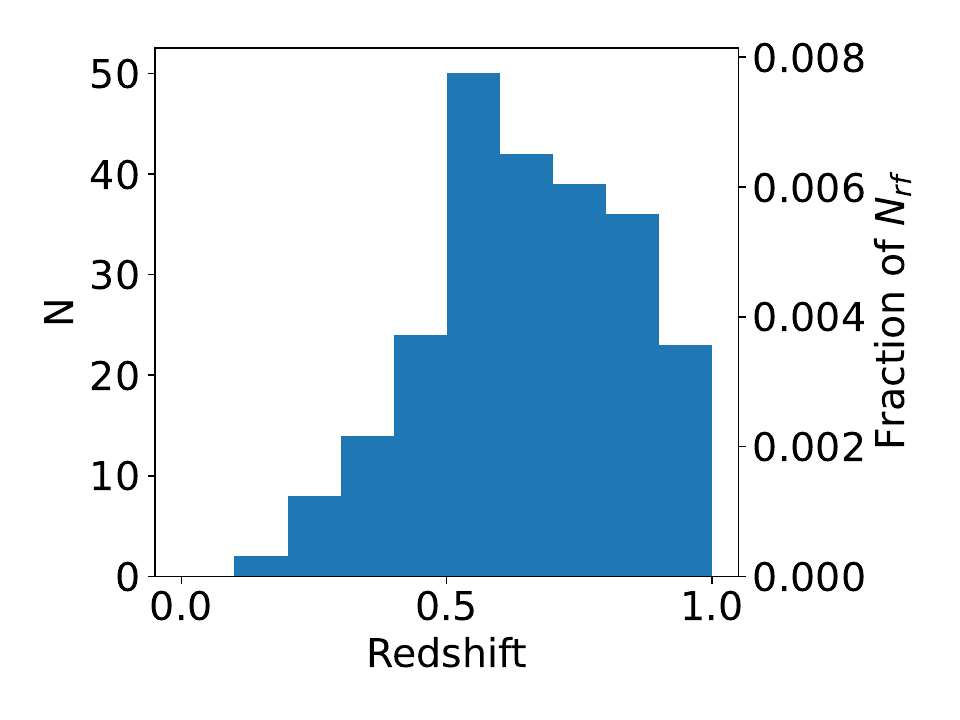}
   \includegraphics[width=0.49\linewidth]{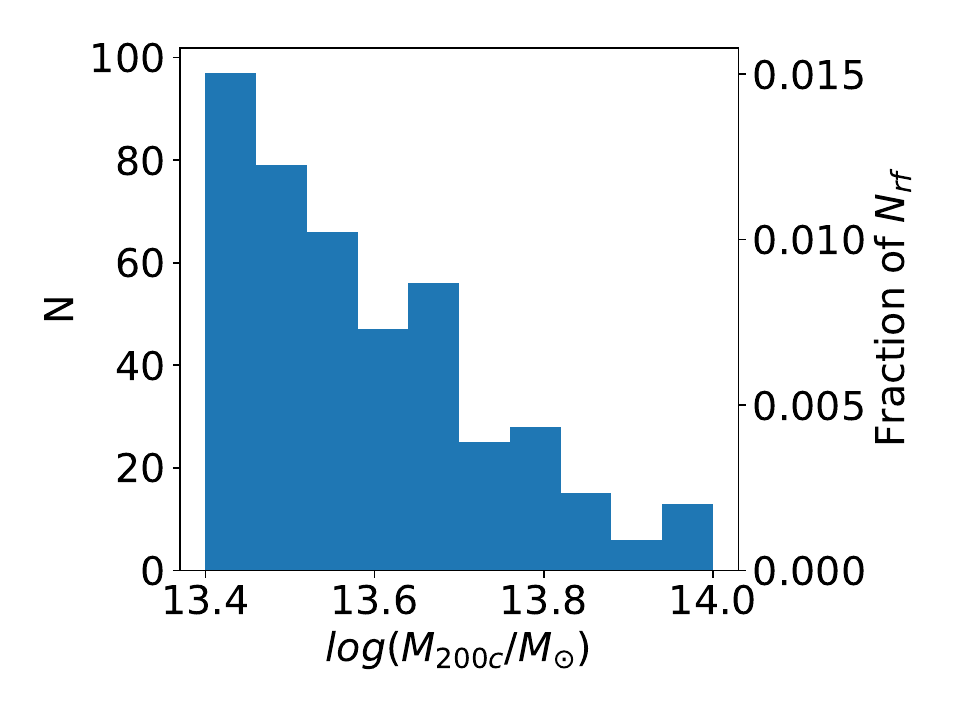}
    \includegraphics[width=0.6\linewidth]{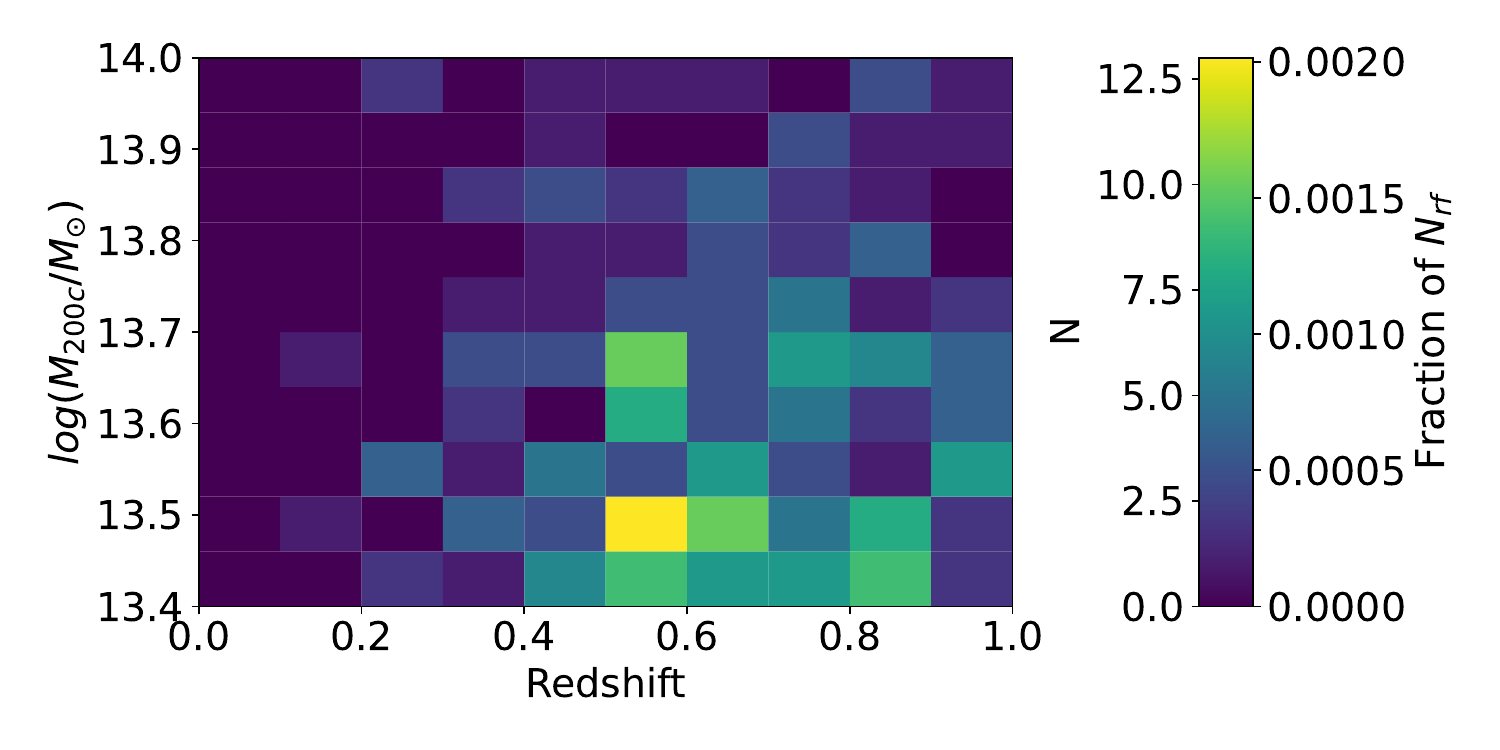}
    \caption{Distribution of the ratio of \YoCL DC2 false positive detections to the total number of random fields $N_{rf}$ as a function of halo mass and redshift (Top) and both (Bottom). In the bottom panel, the scale on the right indicates the number of false positive detections N, and the ratio of N to the total number of random fields $N_{rf}$. The total number of \YoCL random fields is 6,451.
    }
 \label{fig:rnd_zm}
\end{figure*}

\section{Results}
\label{sec:results}

\subsection{Network initial detection catalog}
 We run \YoCL on the training and validation sample for $\sim$100 epochs. Fig.~\ref{fig:loss_lowthr} shows the loss functions for the two samples with different image size. For both cases the training epochs can be split into three parts: 1) in the first epochs the weights converge fast towards optimal values due to the large value of the gradient, 2) the search for an optimal loss minimum (epochs 10-40) and 3) the fine-tuning of the solution. In both cases the lowest value of the validation loss function is reached in the first half of the training epochs -- for 512x512 it was in the range of epochs 10-45, and for 1024x1024 in the range 10-30. 

At each epoch, the network output is a catalog of detections on the validation sample, with the bounding box coordinates, and the probability to belong to the class "cluster" (hereafter detection probability). The network usually outputs multiple detections of the same object, which we discard by following the standard approach in \Yo applications \citep{2015arXiv150602640R,2018arXiv180402767R}. In this case, we define the IoU as the ratio between the area of intersection and the area of union between multiple detection bounding boxes. The gIoU is an optimization of the IoU \citep[see sec. 3.1; ][]{Rezatofighi_2018_CVPR}, and is defined as:
\begin{equation}
    {\rm gIoU} = {\rm IoU} + \frac{\mathcal{U}}{\mathcal{A}_c} - 1 \,
\end{equation}

 where $\mathcal{U}$ and $\mathcal{A}_c$ are the areas of the union of the two boxes and the smallest box enclosing both boxes, respectively.
 
Both the IoU and gIoU are a measurement of the overlap region of bounding boxes that define two different detections.  A value of 1 indicates perfect agreement (we are detecting the same object), while a value approaching 0 indicates increasingly disjointed boxes and/or significantly different sizes (we are detecting different objects).  We discard multiple detections of the same object by applying a gIoU threshold of 0.5, which is the same threshold as in the original \Yo for the IoU \citep{2015A&C....10..121R, 2018arXiv180402767R}. This standard choice means that when two bounding boxes overlap more than 50\%, we consider that they define the same detected object. In this case, we kept the highest probability detection while discarding the other. 

For each epoch, after discarding multiple detections, we obtained a catalog of single detections, each with the coordinates of the bounding box of the detection and the \YoCL probability of the detection being a cluster.

\subsection{Final \YoCL cluster catalog}

 At this point, we needed to choose our best epoch and which probability threshold to use to select the best cluster candidates for our final \YoCL catalog.

 Our best epoch was chosen as the epoch in which the validation loss function reaches its minimum value. This means that in this epoch we reach on average the best values of all the network parameters.

Once we chose the best epoch, to asses our best probability threshold, we used two quantities, the final cluster detection catalog completeness and purity, which are calculated on the \YoCL DC2 detections with respect to our reference DC2 "true cluster" sample from the simulation catalog.  In fact, while we need an hybrid SDSS and DC2 sample for transfer learning, hereafter all our results will focus on the \YoCL performance on DC2 simulations, which are the sample on which we want to test the \YoCL performance on LSST, and which define our cluster catalog selection function.

The cluster catalog completeness quantifies the fraction of true clusters that are detected. The cluster catalog purity quantifies the fraction of detections that are true clusters, as opposite to false positive detections. In machine learning literature, the completeness corresponds to the recall, and the purity to the precision.  
To calculate the purity (see below), we applied \YoCL to a sample of images (“random” fields) that do not contain DC2 clusters, which means that the center of the random fields is more than 12 arcmin ($\sim$4.5~Mpc at $z \gtrsim 0.5$) from any DC2 cluster. For this reason, we added to our validation 6,451 random fields, which correpond to all the regions that do not contain clusters in DC2.

We optimized the detection probability threshold to obtain cluster detection catalogs with the highest values of completeness and purity. Following \citet{2023A&A...677A.101G}, we optimized purity and completeness to the same value, not to have one variable more optimized with respect to the other.  The final \YoCL catalog includes only detections that have a detection probability higher than the optimized detection threshold for which completeness and purity are the same. A more fine-tuned selection function can be defined depending on the use of the catalog for cosmology, galaxy formation and evolution studies, etc.

Fig.~\ref{fig:pc_lowthr} shows the catalog completeness and purity as a function of the detection probability threshold at our best epoch. The completeness $C=\frac{N_{td}}{N_{tc}}$ is calculated as the ratio between the number of true cluster detections $N_{td}$ and the number of true clusters in our images $N_{tc}$. The purity is calculated as $P=1-\frac{N_{fd}}{N_{rf}}$, where $N_{rf}$ is the number of random fields and $N_{fd}$ are cluster detections in the random fields, which are by definition false positive detections.
We assume that the ratio $\frac{N_{fd}}{N_{rf}}$ is a good approximation of the true ratio of false positive detections over the total number of detections, independently on the area of the survey that we consider.
Completeness and purity have the same value of 90\% and 94\% at the threshold value of 27\% and 32\% when using the 512x512 pixel and 1024x1024 pixel images, respectively. 

Figure~\ref{fig:pc_massz} shows the completeness as a function of the DC2 "true cluster" mass M$_{200c}$ and redshift. 
The completeness is almost flat at 0.2<z<0.8 and varies in the range of 80\%-90\%, and 90\%-96\% when \YoCL is applied to 512x512 and 1024x1024 pixel images, respectively. At z>0.8, we observe a decrease in completeness, which is larger when considering 512x512 pixel images. The completeness also increases with the halo mass. For the 512x512 pixel images the completeness is $\gtrsim 95\%$ only for halos with $M_{200c}>10^{14.7}M_{\odot}$, while for 1024x1024 images it is $\gtrsim 94\%$ for $M_{200c}>10^{14}M_{\odot}$.

\subsection{\YoCL final catalog completeness and purity}

Given the higher network  performance with 1024x1024 pixel images, hereafter we concentrate on the catalog obtained with this image size. In this final \YoCL catalog, we only keep cluster candidates with detection probability higher than a 32\% threshold, which corresponds to a catalog 94\% complete and pure. 

Fig.\ref{fig:c_zlam} shows the \YoCL detection catalog completeness as a function of both redshift and DC2 halo  mass and richness. Halo mass and richness are correlated, with a large scatter, and a $M_{200c} = 10^{14}M_{\odot}$ corresponds to a richness $\sim 35$. The \YoCL selection is almost flat with respect to the halo mass up to $z\sim0.9$, but not with respect to richness. This might be due to the fact that the features found by the network to identify a cluster, or the non-linear combination of these features, are more linked with the cluster mass than with its richness.

The catalog is $\sim$100\% complete for $M_{200c} \gtrsim 10^{14.6}M_{\odot}$ and richness $\gtrsim 100$ at all redshifts. At $M_{200c} \gtrsim 10^{14}M_{\odot}$, the completeness is $\gtrsim$95\% up to $z\sim0.8$, and decreases to  $\gtrsim$80-85\% at higher redshifts. However, when characterizing halos by their richness, the completeness is less flat as a function of redshift, as also shown for SDSS observations in \citet{2023A&A...677A.101G}, and  decreases abruptly to $\sim 70-75\%$ at $z>0.8$.

To better understand the purity of \YoCL catalog as function of redshift, we matched the 6\% false detections to lower mass DC2 dark matter haloes, which are the most probable interlopers. Unfortunately, we cannot estimate purity as a function of both mass and redshift because we would need the number of detected clusters with a given observed mass and redshift, and \YoCL does not provide an estimation of these parameters. We found that 49\%, 97\%, and 100\% of the false detections match with halos $10^{13.8} M_{\odot} < M_{200c} < 10^{14} M_{\odot}$, $ 10^{13.5} M_{\odot} < M_{200c}  < 10^{14} M_{\odot}$, and $ 10^{13.4} M_{\odot} < M_{200c}  < 10^{14} M_{\odot}$, respectively,  of which 24\%, 79\%, and 85\% are at z$<1$, respectively.
Fig. \ref{fig:rnd_zm} shows their distributions as a function of mass and redshift. 

Most of the contamination of the final cluster sample is due to groups with $10^{13.7} M_{\odot} < M_{200c} < 10^{14} M_{\odot}$ (i.e., objects with masses within 0.3 dex smaller than a cluster) and $z \gtrsim 0.6$. 

From the DES Y1 redMaPPer cluster catalog~\citep{2019MNRAS.482.1352M}, the cluster mass uncertainty is estimated to be 0.13 dex at $z \lesssim 1$ and $M_{200c}  > 10^{14} M_{\odot}$~\citep{2019MNRAS.490.3341F}. This means that our false positive detections cannot be distinguished from "true clusters" within 3~$\sigma$ of the current DES observational mass uncertainty, which might be taken as a hypothetical lower limit on future LSST cluster mass uncertainties. 

The mass observational uncertainty would also introduce an Eddington bias, which means that the more numerous $M_{200c}  < 10^{14} M_{\odot}$ haloes will be assigned a mass estimation $M_{200c}  > 10^{14} M_{\odot}$, and then contaminate our cluster sample with lower mass groups.  To estimate this bias and using again the current DES cluster mass uncertainty as a reference, we statistically estimated the number of groups in the DC2 footprint with $M_{200c} < 10^{14} M_{\odot}$ that may have a mass estimate of $M_{200c} > 10^{14} M_{\odot}$ due to the scatter of the cluster mass-richness relation, and obtain an Eddington bias of 11\%.

With this hypothesis, this means that $6\%$ of the detections in the \YoCL cluster catalog would be groups with $ 10^{13.4} M_{\odot} < M_{200c}  < 10^{14} M_{\odot}$, and at least $\sim 10\%$ of these groups are expected to be assigned a mass $M_{200c} > 10^{14} M_{\odot}$. In practice, in current surveys the uncertainty on halo mass at $M_{200c} < 10^{14} M_{\odot}$ is about two times larger than the uncertainty at $M_{200c} < 10^{14} M_{\odot}$, about 0.25-0.3 dex \citep[e.g., ][]{2017MNRAS.466.3103S,2017ApJ...848..114P}. If that will be also true for LSST, the Eddington bias contamination will be of the order of $\sim 30\%$, and all these estimates have to be re-assessed when LSST cluster mass uncertainties will be estimated.

\section{Discussion and Conclusions}
\label{sec:discuss}

Our results show that \YoCL detects DC2 clusters ($M_{200c} > 10^{14} M_{\odot}$) in regions centered around them with $\sim 94\%$ completeness and purity at $0.2 \lesssim z \lesssim 1$, and with a 100\% completeness for $M_{200c} > 10^{14.6} M_{\odot}$ within the same redshift range. We also found that the \YoCL selection function is almost flat with respect to the halo mass up to $z\sim0.9$.  In this section, we discuss how this performance compare with other cluster detection methods in optical imaging surveys and other wavelengths.

At lower redshift than LSST, the current DES covers 5,000 sq. deg. in the g, r, i, z and Y bandpasses and reaches a 10~$\sigma$ depth at 24.7, 24.4 and 23.8~mag in g, r and i respectively\footnote{\url{https://des.ncsa.illinois.edu/releases/dr2}}.  This corresponds to a 5~$\sigma$ depth $\sim$~2 mag shallower than LSST. The DES redMaPPer cluster catalog~\citep{2016ApJS..224....1R} is $100\%$ complete for richness $\lambda>70$, which corresponds to a halo mass of $M_{200c}\sim 10^{14.8} M_{\odot}$, using weak lensing and X-ray halo mass estimations for redMaPPer clusters~\citep{2019MNRAS.482.1352M, 2023MNRAS.522.5267U}. Given the large difference in survey depth, it is not surprising that this catalog is less complete than \YoCL DC2 catalog at lower masses.

When comparing to predictions for cluster catalogs completeness and purity at the LSST depth, empirical simulations and a  Bayesian cluster finder~\citep{2012MNRAS.420.1167A,2015MNRAS.453.2515A} predict similar completeness and purity as \YoCL (86-98\%) in the redshift range $0.5 \lesssim z \lesssim 1.0$ for $M_h>10^{14.3} M_{\odot}$ \citep{2017MNRAS.464.2270A}, which corresponds to $M_{200c} > 10^{14.24} M_{\odot}$ \footnote{Hereafter, $M_{200c}$ masses were derived from the original $M_h$ and $M_{500}$ masses found in the literature, using the web-calculator for the equations from \cite{2021MNRAS.500.5056R}, \url{https://c2papcosmosim.uc.lrz.de/static/hydro_mc/webapp/index.html}}.

For observational comparisons, the first survey that reached a depth closer to LSST is the Canada-France-Hawaii Telescope
Legacy Survey (CFHT-LS)\footnote{\url{https://www.cfht.hawaii.edu/Science/CFHLS/}}~\citep{2012AJ....143...38G}. The median 50\% completeness limits in its four deep fields ($\sim 4$~sq. deg.) are 26.3, 26.3 and 25.9 in the g, r, and i bandpasses, respectively~\citep{2007A&A...461..813C}.  When we  analyze the 5~$\sigma$ limit, we obtain similar depths as LSST in these three bands. Several algorithms were applied to the CFHT-LS deep fields to obtain galaxy cluster samples 90-95\% complete at 0.2<z<0.8 and 90\% pure for clusters with richness $\lambda$>50 in simulated data~\citep{2009A&A...494..845G}, and 100\% complete and 85-90 \% pure at $M_{200c}>10^{14.5}M_{\odot}$ ~\citep{2010MNRAS.406..673M}. 

A recent survey that reaches a depth similar to LSST and uses similar optical filters is the Hyper Suprime Camera Strategic Survey Program~\citep[HSC-SSP; ][]{2018PASJ...70S...8A}, which covers an area of $\sim 1,000$ sq. deg. and reaches a 5$\sigma$ depth of 26.8~mag and 26.4~mag in the g and i-bandpasses, respectively. The HSC-SSP cluster catalog obtained with the CAMIRA algorithm ~\citep{2014MNRAS.444..147O} is 100\% and $\sim$ 90\% complete and $ \gtrsim$ 90\%  pure for 
$M_{200c}>10^{14.64}M_{\odot}$ and $M_{200c}>10^{13.94}M_{\odot}$, respectively, in the redshift range 0.1$\lesssim$z$\lesssim$1.1 \citep{2018PASJ...70S..20O}. The CAMIRA algorithm is similar to redMaPPer, and searches for red sequence galaxy overdensities. The WHL09/12 algorithm \citep{2009ApJS..183..197W, 2012ApJS..199...34W},  applied to a compilation of the HSC-SSP and unWISE catalogs,  delivers a cluster catalog 100\% complete for $M_{200c}>10^{14.8} M_{\odot}$ \citep{2021MNRAS.500.1003W},  and 80-90\% complete for $M_{200c}>10^{14.4} M_{\odot}$ at 0.2$\lesssim$z$\lesssim$1. The purity of the sample is not discussed. The completeness significantly decreases for lower cluster mass, reaching $\lesssim$70-60\% completeness for $M_{200c} > 10^{14.1} M_{\odot}$. 

When compared to the completeness and purity expected for Euclid cluster catalogs at $z<1$~\citep{2019A&A...627A..23E} using simulations from \citet{2015MNRAS.453.2515A}, the \YoCL DC2 detections are more complete and pure for $M>10^{14} M_{\odot}$. The best purity and completeness of $\sim$90\% at this mass and redshift ranges were obtained with the algorithm AMICO ~\citep{2018MNRAS.473.5221B}. The other Euclid cluster finder, PZWav, based on wavelet filtering, gives catalogs $\sim$85-87\% complete and pure~\citep{2019A&A...627A..23E}.

Overall, the performance of \YoCL on DC2 simulations is similar or higher when compared to both current optical surveys at the same depth and redshift range, and LSST and Euclid simulation predictions for future cluster catalogs. 

To compare with present and future cluster catalogs obtained at other wavelengths, we compare our results with cluster catalogs obtained by the Sunyaev–Zeldovich ~\citep[SZ; ][]{1972CoASP...4..173S} effect and X-ray flux measurements, which are both sensitive to the cluster hot gas content.

\begin{table*}
    \caption{Completeness of X-ray cluster catalogs. Columns are: 1. Full name of the catalog, 2. Acronym of the catalog, 3. Percentage of the total sky area observed by the survey, 4. Redshift range, 5. Limiting observed flux $F_X$, 6. Completeness for flux $>F_X$, and 7. $M_{200c}$ mass limit at the median redshift, and at z=0.5-1, calculated following \citet{2009A&A...498..361P}. We indicate the redshift (or the redshift range) in parenthesis. }
    \resizebox{!}{8cm}{
    \begin{tabular}{c c c c c c c }
    \hline
    Name & Acronym & Area & Redshift range & $F_X$ & Comp. & $M_{200c}$ limit \\
          &  & \% of sky &   & $erg/cm^2/s$ & \% & $M_{\odot}$\\
    \hline
    & &   & & & & \\
       {\bf MCXC: ROSAT }& &   & & & & \\
   & &   & & & & \\
   ROSAT-ESO Flux-Limited X-Ray & REFLEX$^a$ & 33 & 0<z<0.3 & $3 \cdot 10^{-12}$ & 100 & $10^{13.9}$ (0.075) \\
    & &   & & & & $10^{15.1-15.6}$ (0.5-1)\\
    & &   & & & & \\
    Northern ROSAT All-Sky & NORAS$^b$ & 41 & 0<z<0.3 & $3 \cdot 10^{-12}$ & 50 & $10^{13.9}$ (0.075)\\
   Galaxy Cluster Survey & &   & & & & $10^{15.1-15.6}$ (0.5-1)\\
        & &   & & & & \\
   ROSAT Brightest Cluster Sample& BCS$^c$ & 41 & 0<z<0.3 & $4.4 \cdot 10^{-12}$ & 90 & $10^{13.9}$ (0.075)\\
   & &   & & & & $10^{15.1-15.6}$ (0.5-1)\\
        & &   & & & & \\
    Catalog of clusters in the region of & SGP$^d$ & 8  & 0<z<0.3 & $3  \cdot 10^{-12}$  & 100 & $10^{13.9}$ (0.075)\\
 1 ster. around the south galactic pole &  &   &  &  & & $10^{15.1-15.6}$ (0.5-1)\\
        & &   & & & & \\
   ROSAT north ecliptic pole survey & NEP$^e$ & 0.2  & 0<z<0.8 & $2.0 \cdot 10^{-14}$ & 100 & $10^{13.9}$ (0.200)\\
       & &   & & & & $10^{14.5-15.0}$ (0.5-1)\\
            & &   & & & & \\
Massive Cluster Survey & MACS$^f$ & 55  & 0.3<z<0.6 & $2 \cdot 10^{-12}$ &  93 & $10^{13.9}$ (0.370)\\
    & &   & & & & $10^{14.5-15.0}$ (0.5-1)\\
                   &      &     &           & $ <2  \cdot 10^{-12}$ &  59 & \\
      &      &     &           &  &   &\\
{ \bf MCXC: Serendipitous surveys} & &   & & & & \\
& &   & & & & \\
160 Square Degree ROSAT Survey & 160SD$^g$ & 3.8 & 0<z<0.7 & $1.4 \cdot 10^{-14}$ & 100 & $10^{13.6}$ (0.250) \\
          & &   & & & & $10^{14.3-14.8}$ (0.5-1)\\
            & &   & & & & \\
  400 Square Degree ROSAT PSPC & 400SD$^h$ & 9.5 & 0<z<0.7 & $1.4 \cdot 10^{-14}$ & 100 & $10^{13.6}$ (0.200)\\
   Galaxy Cluster Survey &  &   &  &  & &$10^{14.5-15.0}$ (0.5-1)\\
               & &   & & & & \\
  Southern Serendipitous High-redshift & SHARC$^i$ & 0.4 & 0<z<0.7 & $4.6 \cdot 10^{-14}$ & 100 & $10^{13.6}$ (0.340) \\
  Archival ROSAT Cluster survey &  &   &  &  & & $10^{14.5-15.0}$ (0.5-1)\\
               & &   & & & & \\
  Extended Medium-Sensitivity Survey & EMSS$^j$ & 1.8 & 0<z<0.7 & $5  \cdot 10^{-14}..3 \cdot 10^{-12}$ & 100 & $10^{13.6}$ (0.115) \\
  Distant Cluster Sample &  &   &  &  & & $10^{14.5-15.0}$ (0.5-1)\\
               & &   & & & & \\
  Wide Angle ROSAT Pointed Survey & WARPS$^k$ & 1.3 & 0<z<0.9 & $6.5 \cdot 10^{-14}$ & 100 & $10^{13.6}$ (0.284) \\
  Distant Cluster Sample &  &   &  &  & & $10^{14.2-14.7}$ (0.5-1)\\
               & &   & & & & \\
 {\bf eROSITA Final Equatorial-Depth Survey} & eFEDS$^l$ & 3.3 & 0<z<1.3 & $1 \cdot 10^{-14}$ & 40 & $10^{13.8}$ (0.353) \\
     &  &   &  &  & & $10^{14.1-14.6}$ (0.5-1)\\
               & &   & & & & \\
   {\bf eRASS1 cosmology cluster sample}$^m$ &  & 50 & 0<z<1.3 & $1.6 \cdot 10^{-12}$ & 80 & $10^{14.4}$ ($\sim0.300$) \\
       &  &   &  &  & & $10^{14.1-14.6}$ (0.5-1)\\
    \hline
     \end{tabular}
     }\\
    \footnotesize{$^a$ \citet{2004A&A...425..367B}; $^b$\citet{2000ApJS..129..435B}; $^c$\citet{1998MNRAS.301..881E}; $^d$\citet{2002ApJS..140..239C}; $^e$\citet{2006ApJS..162..304H}; $^f$\citet{2001ApJ...553..668E}; $^g$\citet{2003ApJ...594..154M}; $^h$\citet{2007ApJS..172..561B}; $^i$\citet{2000ApJS..126..209R, 2003MNRAS.341.1093B}; $^j$\citet{1990ApJ...356L..35G}; $^k$\citet{2002ApJS..140..265P, 2008ApJS..176..374H};$^l$\citet{2022A&A...661A...1B}; $^m$\citet{2024A&A...682A..34M, 2024arXiv240208453K} }
    \label{tab:mcxc}
\end{table*}

SZ cluster catalogs are mass-limited and the deepest catalogs available at present reach 100\% completeness at $M_{200c} > 10^{14.86}-10^{14.94}M_{\odot}$ at $z \lesssim 1.5$ from observations with the South Pole Telescope Polarimeter ~\citep[SPTPol; ][]{2020ApJS..247...25B}, a much higher mass limit than optical and infrared surveys. The SPT-SZ survey~\citep{2015ApJS..216...27B} catalog is 100\% complete at $M_{200c} >  10^{14.94}-10^{15.00}M_{\odot}$ in a similar redshift range. The cluster catalog obtained from the fifth data release (DR5) of observations (13,211 deg$^2$) with the Atacama Cosmology Telescope (ACT) is 90\% complete for the clusters with $M_{200c} > 10^{14.76-14.66}$ at 0.2<z<2.0~\citep{2021ApJS..253....3H}. The Planck space mission PSZ2 all-sky cluster catalog~\citep{2016A&A...594A..27P} is 80\% complete for $M_{200c} > 10^{14.76} M_{\odot}$ at 0.4$<$z$<$0.6, and for $M_{200c} > 10^{14.3} M_{\odot}$ for clusters at z$\sim$0.2. 

Simulations of the current SPT-3G survey, which will provide much deeper observations~\citep{2014SPIE.9153E..1PB}, were used to estimate the completeness and purity that can be attained with another deep convolutional neural network~\citep{2021MNRAS.507.4149L}, combined with a classical match filter~\citep{2006A&A...459..341M}. This work shows that $\sim$95\% completeness and purity is predicted to be attained at $M_{200c} > 10^{14.7} M_{\odot}$ at $z \gtrsim 0.25$. 

This means that all present SZ surveys reach $\sim 95\%$ completeness at cluster masses much higher than what is predicted for LSST from this work. However, the next generation SZ experiments, like SPT-3G, Simons Observatory, CMB-S4 will obtain cluster catalogs with a limiting mass $M_{200c} \sim 10^{14} M_{\odot}$ more comparable to the LSST mass limit~\citep{2022ApJ...928...16R}. The CMB-S4 WIDE~\citep{2016arXiv161002743A} survey will reach the S/N=5 cluster detection limit of  $M_{200c} = 10^{14.1} M_{\odot}$ at the redshift range 0.2<z<1 over 67\% of the sky; the S/N=5 detection threshold for the Simons Observatory~\citep{2019JCAP...02..056A} is planned to be $M_{200c} = 10^{14.3} M_{\odot}$ in the same redshift range over 40\% of the sky; and the CMB-S4 ULTRADEEP and CMB-HD~\citep{2019BAAS...51g...6S} surveys are built to reach up to $M_{200c} = 10^{14} M_{\odot}$ and $M_{200c} = 10^{13.8} M_{\odot}$, respectively. However, the CMB-S4 ULTRADEEP survey covers only 3\% of the sky, while CMB-HD is planned to cover $\sim$ 50\% of the sky. All these survey are planned for $ \gtrsim 2030$, most probably about at the same time as the the 5-year LSST data release.


For what concerns X-ray surveys, the reference X-ray all-sky cluster catalog is the Röntgensatellit~\citep[ROSAT; ][]{1987SPIE..733..519P, 1999A&A...349..389V} catalog of Extended Brightest Cluster Sample~\citep[BCS; ][]{1998MNRAS.301..881E}, which contains 201 cluster in Northern hemisphere and is 90\% complete for $z<0.3$ and X-ray fluxes higher than 4.4 $\cdot 10^{-12}$ erg/cm$^2$/s. 
The MCXC cluster catalog~\citep{2011A&A...534A.109P} is a compilation of several catalogs/surveys that consists of ROSAT-based catalogs and serendipitous catalogues, summarized in Table~\ref{tab:mcxc}. As expected, X-ray surveys detect clusters at much higher masses than LSST at z=0.5-1.

The ComPRASS catalog~\citep{2019A&A...626A...7T} presents a compilation of Planck~\citep{2016A&A...594A..27P} and RASS~\citep{2004A&A...423..449P} catalogs of galaxy clusters that were observed in X-ray and using SZ, and reaches deeper than each survey used to compile it. Therefore, the selection function is a complicated combination of the selection function of several surveys. CompRASS is 100\%  complete for $M_{200c} > 10^{14.6}M_{\odot}$, $M_{200c} > 10^{14.8}M_{\odot}$, ans $M_{200c} > 10^{14.7}M_{\odot}$ at z$<$0.3 and z$<$0.6,  and 0.6$<$z$<$1.0, respectively, which are much lower than the completeness limit for the SZ catalogs used to build it. 


In conclusion, \YoCL shows similar completeness and purity as other algorithms applied to current deep optical imaging surveys like CFHTLS Deep and HSC-SSP, and better completeness and purity than most of the other methods that have been applied to Euclid simulations.
Compared to current SZ and X-ray surveys, \YoCL can obtain more complete and pure catalogs at much lower masses. However, future SZ surveys are planned to provide much deeper complete and pure catalogs directly comparable with ours. With respect to this, we notice that both SZ surveys and the \YoCL selection function are mass-limited, making the SZ-optical comparison based on similar selection functions. \YoCL detections can also be combined to SZ and X--ray detections as it was done for the ComPRASS compilation, to reach catalogs with higher completeness and purity at lower masses.

It has to be noticed that in this paper we focus our analysis on the targeted detections, with the goal to analyze the performance of the algorythm itself, independently of possible systematics and biases  introduced by the variations of the parameters of the images generated in a survey mode. In future papers, we will apply \YoCL to DC2 images in a survey mode, and our detections will be compared to other LSST cluster detection algorithms applied to the DC2 simulations.

\label{sec:Train}

\section{Summary}
\label{sec:summary}

We applied the \YoCL deep convolutional network \citep{2023A&A...677A.101G} to observations from SDSS and DESC DC2 simulations to estimate its performance for LSST.  We trained the network on 12,203 and 1,171 g, r and i composite color images from SDSS and from the DESC DC2 simulations, respectively, and validated on the same number of cluster images (for a total of 24,406 SDSS and 2,342 DC2 training and validation images) and 6,451 random fields.  We conclude that:



\vspace{2mm}

\begin{itemize}

\item When using DC2 LSST simulated images with a pixel size equal to four times the LSST pixel resolution ($\approx$ 0.8''/pix), the \YoCL  DC2 cluster catalog is 94\% pure and complete for $M_{200c} > 10^{14} M_{\odot}$ and at $0.2<z<1$, and 100\% complete for $M_{200c} > 10^{14.6} M_{\odot}$. 

\vspace{2mm}

\item The cluster selection function is mass-limited at 0.2<z<0.9. 

\vspace{2mm}

\item When compared to other cluster detection methods in current optical surveys that reach LSST depth and simulations of the Euclid surveys, \YoCL shows similar or better completeness and purity. 

\vspace{2mm}

\item Current X-ray and SZ cluster surveys do not reach \YoCL completeness and purity at $M_{200c} > 10^{14} M_{\odot}$ and at $0.2<z<1$, while future SZ surveys will be directly comparable to LSST \YoCL detections and will have similar mass-limited selection functions.
\end{itemize}
\vspace{2mm}
This paper shows that \YoCL will permit us to obtain LSST cluster catalogs that will be 94\% pure and complete for $M_{200c} > 10^{14} M_{\odot}$ and at $0.2<z<1$, and 100\% for $M_{200c} > 10^{14.6} M_{\odot}$ The \YoCL cluster selection function is mass-limited in the redshift range 0.2<z<0.9. We focused our analysis on targeted detections, with the goal to analyze the performance of the algorythm itself, independently of possible systematics and biases  introduced by a survey mode. 

We compare our algorithm to other cluster detection methods in current optical surveys that reach LSST depth and simulations of the Euclid surveys, and \YoCL shows similar or better completeness and purity. When compared to current X-ray and SZ cluster surveys \YoCL reaches higher completeness and purity at $M_{200c} > 10^{14} M_{\odot}$ and at $0.2<z<1$. However, future SZ surveys will reach similar completeness and purity at the same depth as LSST \YoCL detections, and will have similar mass-limited selection functions.

We note that this analysis was based on LSST DC2 images and did not involve the image processing required to obtain galaxy photometric and photometric redshift catalogs, or the masking of stellar sources and artifacts. The advantage of this deep machine learning approach that works directly on images is to obtain cluster catalogs that will be complementary to other optical detection methods used in the LSST DESC collaboration, and that will be independent from systematic and statistical uncertainties inherent to galaxy catalog production.

In future papers, we will study the \YoCL performance in survey mode, and our detections will be compared to other LSST cluster detection algorithms.

\label{sec:Conclusion}

\begin{acknowledgements}
We thank Universit\'e Paris Cité (UPC), which founded KG's Ph.D. research. We gratefully acknowledge support from the CNRS/IN2P3 Computing Center (Lyon - France) for providing computing and data-processing resources needed for this work. 
We describe below the author’s contributions. Kirill Grishin applied YOLO-CL to the DC2 simulations, produced the results and figures in the paper, and was the main writer of Sections 2.2 and 5. Simona Mei co-conceived the YOLO-CL network with Stéphane Ilic, developed the content of this paper, supervised the work of Kirill Grishin, Stéphane Ilic and Michel Aguena, and was the main writer of the paper's text, answered the internal DESC reports. She is the contact with the editor. Stéphane Ilic modified the original YOLO network to adapt it for galaxy cluster detection. He co-conceived YOLO-CL with Simona Mei and developed the network and analysis software to derive the completeness and purity plots. Michel Aguena contributed to the generation and validation of the DC2 images, and to the analysis and discussion of the cluster detection, including the improvement on the purity estimation model. He also shaped the final image generation software used, and provided the masses and richnesses estimations to the dark matter halo catalog. Dominique Boutigny and Marie Paturel helped with image generation at the beginning of the project and experimented with different versions of YOLO. These statements have been validated with the DESC publication board after having the confirmation of the authors. 
The Dark Energy Science Collaboration (DESC) acknowledges ongoing support from the IN2P3 (France), the STFC  (United Kingdom), and the DOE, NSF, and LSST Corporation (United States). As members of the DESC collaboration, we used resources of the IN2P3 Computing Center  (CC-IN2P3--Lyon/Villeurbanne - France) funded by the Centre National de la Recherche Scientifique; the National Energy Research Scientific Computing Center, a DOE Office of Science User Facility supported under Contract  No.\ DE-AC02-05CH11231; STFC DiRAC HPC Facilities, funded by UK BEIS National  E-infrastructure capital grants; and the UK particle physics grid, supported by the GridPP Collaboration.  This work was performed in part under DOE  Contract DE-AC02-76SF00515. This paper has undergone an internal review by the LSST DESC, and we thank the internal reviewers, Camille Avestruz and Markus Michael Rau, for fruitful discussions that improved the paper.
\end{acknowledgements} 
\clearpage

\bibliographystyle{aa}
\bibliography{References.bib}

\end{document}